\documentclass[3p,times]{elsarticle}
\usepackage[colorlinks=true]{hyperref}
\usepackage{amssymb}
\biboptions{authoryear}

\usepackage[figuresright]{rotating}

\usepackage{color}

\newcommand{\Ham}{\mathcal{H}}     
\newcommand{\spin}[1]{\sigma_{#1}} 
 
\newcommand{\plaq}[4]{\left[#1,#2,#3,#4\right]} 
 
\newcommand{\plaqsum}{
  \sum\limits_{\plaq{i}{j}{k}{l}}\spin{i}\spin{j}\spin{k}\spin{l}}

\begin{document}

\title{Transmuted finite-size scaling at first-order phase transitions}

\author[Leipzig]{Marco Mueller}
\ead{Marco.Mueller@itp.uni-leipzig.de}

\author[Leipzig]{Wolfhard Janke}
\ead{Wolfhard.Janke@itp.uni-leipzig.de}

\author[HW]{Desmond A.\ Johnston}
\ead{D.A.Johnston@hw.ac.uk}

\address[Leipzig]{Institut f\"ur Theoretische Physik, Universit\"at Leipzig,
   Postfach 100\,920, 04009 Leipzig, Germany}
   
\address[HW]{Department of Mathematics, 
School of Mathematical and Computer Sciences,
Heriot-Watt University, Riccarton, Edinburgh EH14 4AS, Scotland}

\begin{abstract}

It is known that fixed boundary conditions modify the leading finite-size corrections for an $L^3$ lattice in $3d$  at a first-order phase transition from $1/L^3$ to $1/L$. We note that an exponential low-temperature phase degeneracy of the form $2^{3L}$ will lead to a different leading correction of order $1/L^2$. A $3d$ 
gonihedric Ising model with a four-spin interaction, plaquette Hamiltonian displays such a degeneracy
and we confirm the modified scaling behaviour using  high-precision
multicanonical simulations. 

We remark that other models such as the Ising antiferromagnet on the FCC lattice, in which the number of  ``true'' 
low-temperature phases
is not 
macroscopically large
but which possess an exponentially degenerate number of low lying states may display an effective version of the modified scaling law for the range of lattice sizes accessible in simulations.\\
\end{abstract}

\maketitle

\vspace*{8pt}

To some extent first-order phase transitions have been the poor cousins of continuous transitions when it comes to numerical investigations, in spite of their prevalence in nature, see, e.g., the articles in \cite{1st-order-book}.
Initial studies of finite-size scaling for first-order transitions were carried out by \cite{pioneer1,pioneer2,pioneer3}
and further pursued by \cite{furtherd1,furtherd2,furtherd3,furtherd4,furtherd5}. Later, rigorous results  for periodic boundary conditions were  derived using Pirogov-Sinai theory and similar techniques applied to the case of non-periodic boundary conditions, for a review see \cite{janke_review}. 

It is  possible to derive the finite-size scaling behaviour at first-order phase transitions
using straightforward heuristic arguments discussed in \cite{twophasemodel1}, rather than the more sophisticated approach of \cite{rigorous1,rigorous2}. To do this, we introduce a simple two-phase model in which the
system spends a fraction $W_{\rm o}$ of its total time in one of  $q$
ordered phases (as in a $q$-state Potts model) and a fraction $W_{\rm d} = 1 - W_{\rm o}$ in the disordered
phase.  The corresponding energies are $\hat{e}_{\rm o}$ and $\hat{e}_{\rm d}$,
respectively, where the hat is introduced for quantities evaluated at the inverse
phase transition temperature of the infinite system, $\beta^\infty$.
Physically, the model neglects fluctuations within the phases and also treats 
the flips between the phases
as instantaneous jumps. With these assumptions the energy moments are just the weighted average
over the phases,
\mbox{$\left<e^n\right> =
W_{\rm o}\hat{e}_{\rm o}^n + (1-W_{\rm o})\hat{e}_{\rm d}^n$},
and from this we can calculate various other derived observables.
The specific heat
$C_V(\beta, L) = -\beta^2\partial e(\beta, L)/\partial\beta$ is given by
\begin{equation}
  C_V(\beta, L) = L^d\beta^2\left(\left< e^2\right> - \left< e \right>^2\right) =
  L^d\beta^2 W_{\rm o}(1-W_{\rm o})\Delta \hat{e}^2
  \label{eq:specheat}
\end{equation}
with $\Delta \hat{e} = \hat{e}_{\rm d} - \hat{e}_{\rm o}$. Differentiating with respect to $W_{\rm o}$ then gives a maximum
$C_V^{\rm max} = L^d (\beta^\infty\Delta \hat{e}/2)^2$ 
at $\beta^{C_V^{\rm max}}(L)$ for $W_{\rm o} = W_{\rm d}
= 0.5$, where the disordered and
ordered peaks of the energy probability density have equal {\it weight}. We can see that this immediately recovers the $L^d$ scaling of the peak in  $C_V^{\rm max}$.

The leading corrections can also be obtained by Taylor expanding the ratio of weights around $\beta^\infty$.
The probability of being in any one of the ordered states or the disordered state is
related to the free-energy densities $\hat{f}_{\rm o}, \hat{f}_{\rm d}$ of the states by
\begin{equation}   
p_{\rm o}\propto e^{-\beta L^d \hat{f}_{\rm o}}\mbox{ and } p_{\rm d} \propto
  e^{-\beta L^d \hat{f}_{\rm d}}
\end{equation}
and the fraction of time spent in the ordered states is
proportional to $q p_{\rm o}$. The ratio of weights is thus $W_{\rm
o}/W_{\rm d} \simeq q e^{- L^d \beta \hat{f}_{\rm o}}/ e^{-\beta L^d \hat{f}_{\rm d}}$ (up to exponentially small corrections in $L$, see \cite{borgs-janke,twophasemodel1}).
Taking the logarithm of this ratio gives $\ln (W_{\rm o}/W_{\rm d}) \simeq \ln q + L^d\beta(\hat{f}_{\rm d}
- \hat{f}_{\rm o})$ and expanding  around $\beta^\infty$ at the finite-size specific-heat maximum where $W_{\rm o} = W_{\rm d}$ gives
   $0 = \ln q + L^d\Delta \hat{e}(\beta -
  \beta^\infty) + \dots$,
which can be solved for the finite-size peak loation of the specific heat:
\begin{equation}
  \beta^{C_V^{\rm max}}(L) = \beta^\infty - \frac{\ln q}{L^{d}\Delta\hat{e}} +
  \dots \; .
  \label{eq:fss:beta:specheat}
\end{equation}
The $1/ L^d$ leading correction to scaling is immediately apparent.
Similar calculations of~\cite{twophasemodel1}
for the location $\beta^{B^{\rm min}}(L)$ of the minimum of the energetic
Binder parameter 
\begin{equation}
  B(\beta, L) = 1 - \frac{\langle e^4\rangle}{3\langle e^2 \rangle^2} 
  \label{eq:binder}
\end{equation}
give
\begin{equation}
  \beta^{B^{\rm min}}(L) = \beta^\infty - \frac{\ln(q\hat{e}^2_{\rm
  o}/\hat{e}^2_{\rm d})}{L^d\Delta \hat{e}}  + \dots 
  \label{eq:fss:beta:binder}
\end{equation}
which again displays the $1/ L^d$ correction.

\begin{figure}[b]
    \centering
    \includegraphics[width=2.5cm]{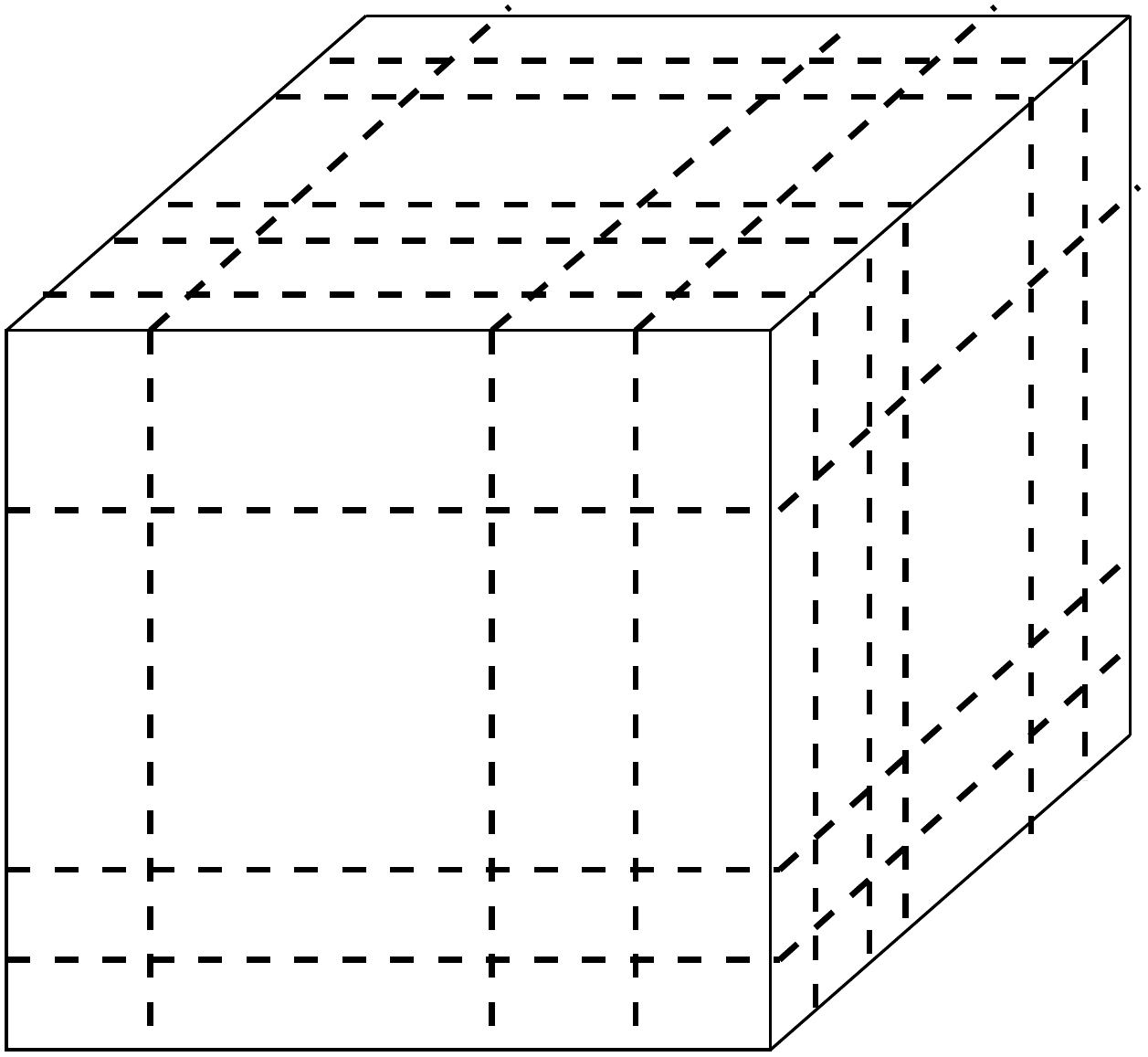}
    \caption{A typical ground state of the $3d$ plaquette Hamiltonian showing  planes of spins flipped with respect to a purely ferromagnetic ground state dotted. Since {\it any} plane of spins may be flipped, the degeneracy is $q=2^{3L}$.} 
    \label{fig:sketch}
  \end{figure}

The key observation is now that an exponential degeneracy in $q$, the number of low-temperature phases, will
alter the  scaling behaviour because of the presence of the various $\ln(q)$ factors in the leading scaling terms in Eqs.~(\ref{eq:fss:beta:specheat}), (\ref{eq:fss:beta:binder}). 
One  model with precisely this feature is a $3d$ plaquette (4-spin) interaction Ising model on a cubic lattice, 
\begin{equation}
  \Ham = -\frac{1}{2}\plaqsum \;,
  \label{eq:ham:gonikappa0}
\end{equation}
where the ground-state degeneracy of $q=2^{3L}$ on an $L^3$ lattice (cf.\ Fig.\ \ref{fig:sketch})
was shown by \cite{degen} to be unbroken throughout the low-temperature phase.
This is a member of a  family of so-called gonihedric Ising models 
which were originally formulated as a lattice discretization of string-theory actions in {high-energy} physics  which depend solely on the extrinsic curvature of the string worldsheet, for a review see \cite{goni_review}.
This plaquette Hamiltonian has 
attracted  attention because it displays a strong first-order transition observed first by
\cite{firstorder}
and evidence of glass-like behaviour at low temperatures discovered by \cite{glassy1,goni_review}.

In the plaquette gonihedric model with $q=2^{3L}$, Eqs.~(\ref{eq:fss:beta:specheat}), (\ref{eq:fss:beta:binder})
become
\begin{equation} 
	\beta^{C_V^{\rm max}}(L) 
    = \beta^\infty - \frac{\ln
    2^{3L}}{L^3\Delta\hat{e}} 
    + {\cal O}\left( (\ln 2^{3L})^2 L^{-6} \right)
     = \beta^\infty - \frac{3\ln 2}{L^{2}\Delta\hat{e}} + {\cal O}\left( L^{-4}
    \right)
    \label{eq:fss:beta:specheat2} \\
\end{equation} 
and
\begin{equation} 
  \beta^{B^{\rm min}}(L) 
  = \beta^\infty - \frac{\ln(2^{3L}\hat{e}^2_{\rm
  o}/\hat{e}^2_{\rm d})}{L^3\Delta \hat{e}}
  + {\cal O}\left(
  (\ln(2^{3L}\hat{e}^2_{\rm o}/\hat{e}^2_{\rm d}))^2 L^{-6}\right)
   = \beta^\infty - \frac{3\ln2}{L^2\Delta \hat{e}}  - \frac{\ln(\hat{e}^2_{\rm
  o}/\hat{e}^2_{\rm d})}{L^3\Delta\hat{e}} + {\cal O}\left( L^{-4}\right)
  \label{eq:fss:beta:binder2}
\end{equation} 
so the leading contribution to the finite-size corrections is now, as expected with the exponential degeneracy,  $\propto
L^{-2}$.  For the extremal values one also finds non-standard scaling corrections 
\begin{equation} 
  C_V^{\rm max}(L) = L^3\left( \frac{\beta^\infty \Delta\hat{e}}{2}\right)^2 +
  {\cal O}(L)
  \label{eq:fss:specheat} 
\end{equation} 
and 
\begin{equation} B^{\rm min}(L) = 1 - \frac{1}{12}\left( \frac{\hat{e_{\rm o}}}{\hat{e_{\rm d}}} +
  \frac{\hat{e_{\rm d}}}{\hat{e_{\rm o}}} \right)^2  + {\cal O}(L^{-2}) \; ,
  \label{eq:fss:binder} 
\end{equation}
where the leading correction terms are also modified by a factor of $L$ due to the exponential degeneracy compared with the standard case.

To verify the modified scaling
we used in \cite{highero} the multicanonical
Monte Carlo algorithm of \cite{muca1,muca2,mucaweights1,mucaweights2} where rare states lying between the ordered and disordered phases are  promoted
artificially, decreasing the autocorrelation time and allowing the system to
oscillate more rapidly between phases. 
We systematically improve guesses of the energy
probability distribution 
using recursive estimates of \cite{mucarecursion} before the actual 
production run with of the order of $(100 - 1000) \times 10^6$ sweeps.
Canonical estimators can then be
retrieved by weighting the multicanonical data to yield Boltzmann-distributed
energies. Reweighting techniques are very powerful when combined with multicanonical
simulations, and allow the calculation of observables over a broad range of
temperatures.
Errors on the measured quantities have been extracted by jackknife 
analysis using $20$
blocks for each lattice size.
The observables such as the specific heat (\ref{eq:specheat}) and
Binder's energy parameter (\ref{eq:binder}) have been calculated 
{from the data as function of temperature by reweighting}.
This enables us to determine the
positions of their peaks, $\beta^{C_V^{\rm max}}(L)$ and $\beta^{B^{\rm
min}}(L)$, 
with high precision.

\begin{figure}[h] 
	\begin{center} (a)\hspace{-4ex}
		\includegraphics[scale=0.65]{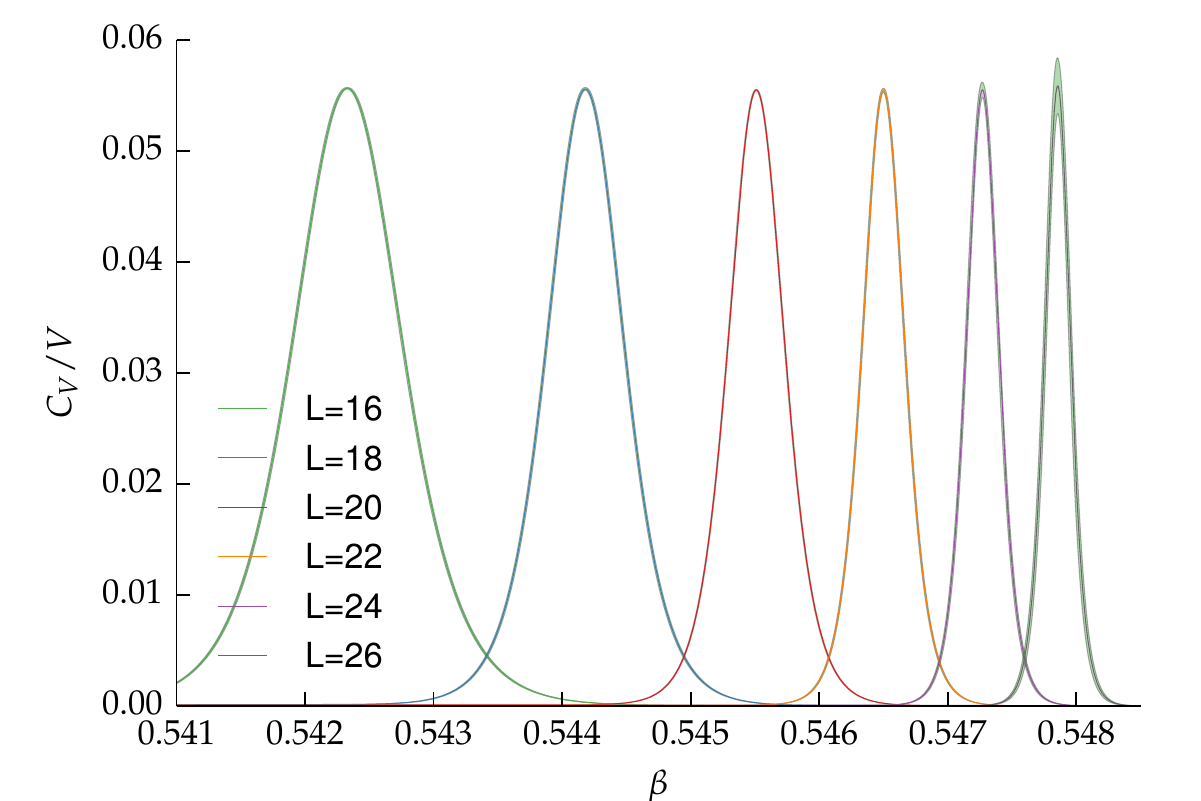}\hfill (b)\hspace{-2ex}\includegraphics[scale=0.65]{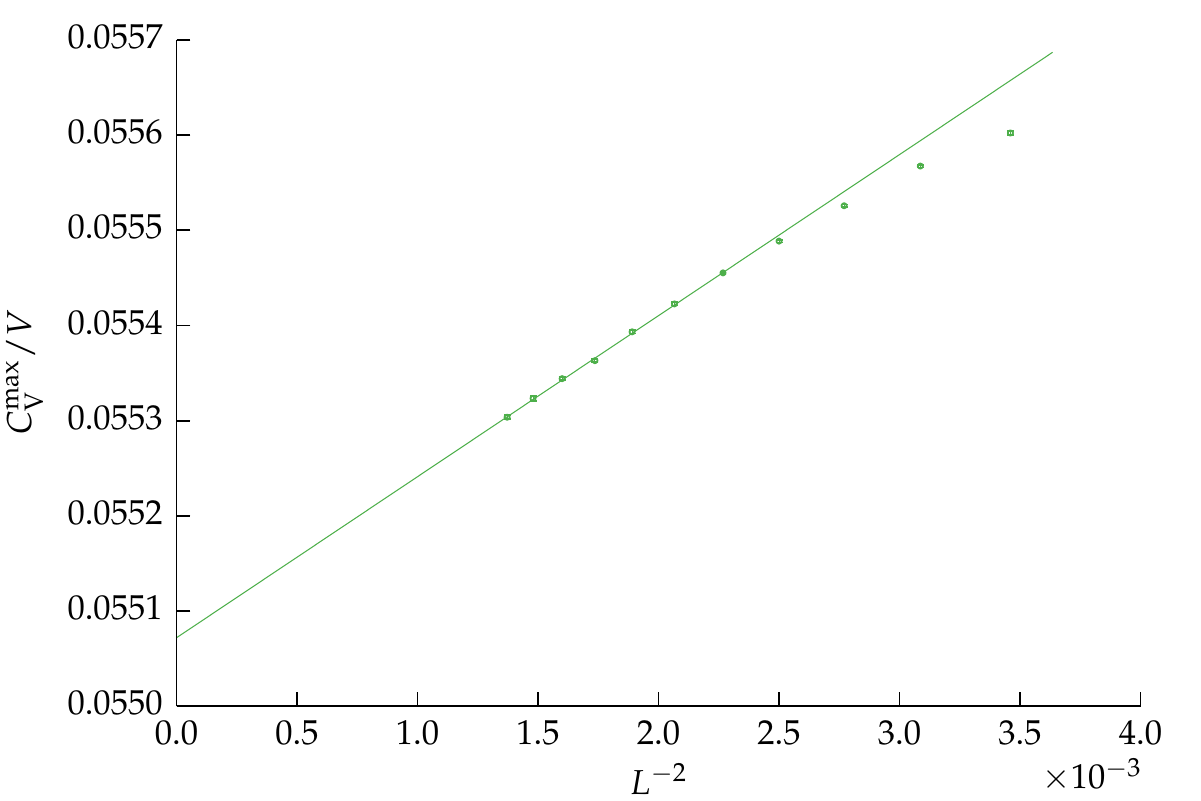} 
		\caption{(a) Specific-heat curves as function of $\beta$ and (b) their maxima, $C_V^{\rm max}(L)/V$, vs $1/L^2$ showing clearly the (non-standard) ${\cal O} (L^{-2})$ behaviour of the leading correction. The best fit line of $0.055072(4) + 0.1693(21) L^{-2}$ through the measured values is drawn.}
    \label{fig:cv_max_fit_nt} 
  \end{center} 
\end{figure} 

\begin{figure}[h] 
	\begin{center} (a)\hspace{-4ex}
		\includegraphics[scale=0.65]{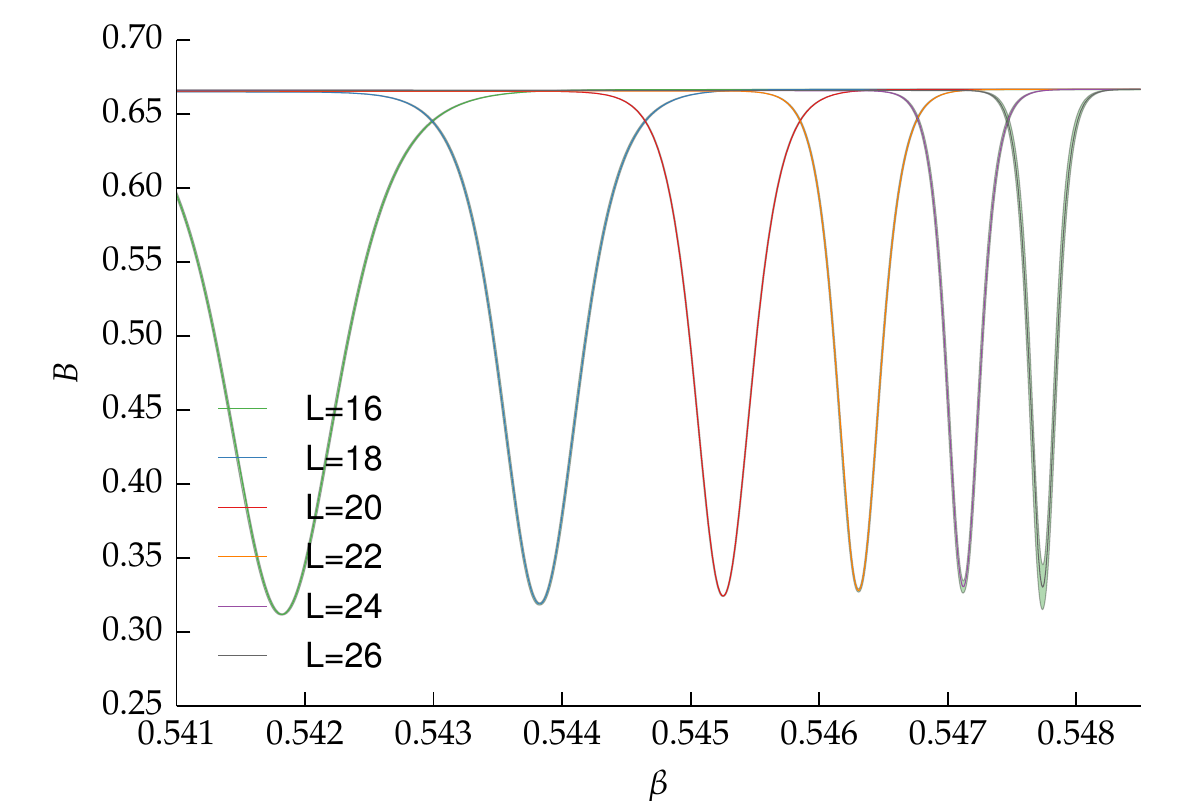}\hfill(b)\hspace{-2ex}\includegraphics[scale=0.65]{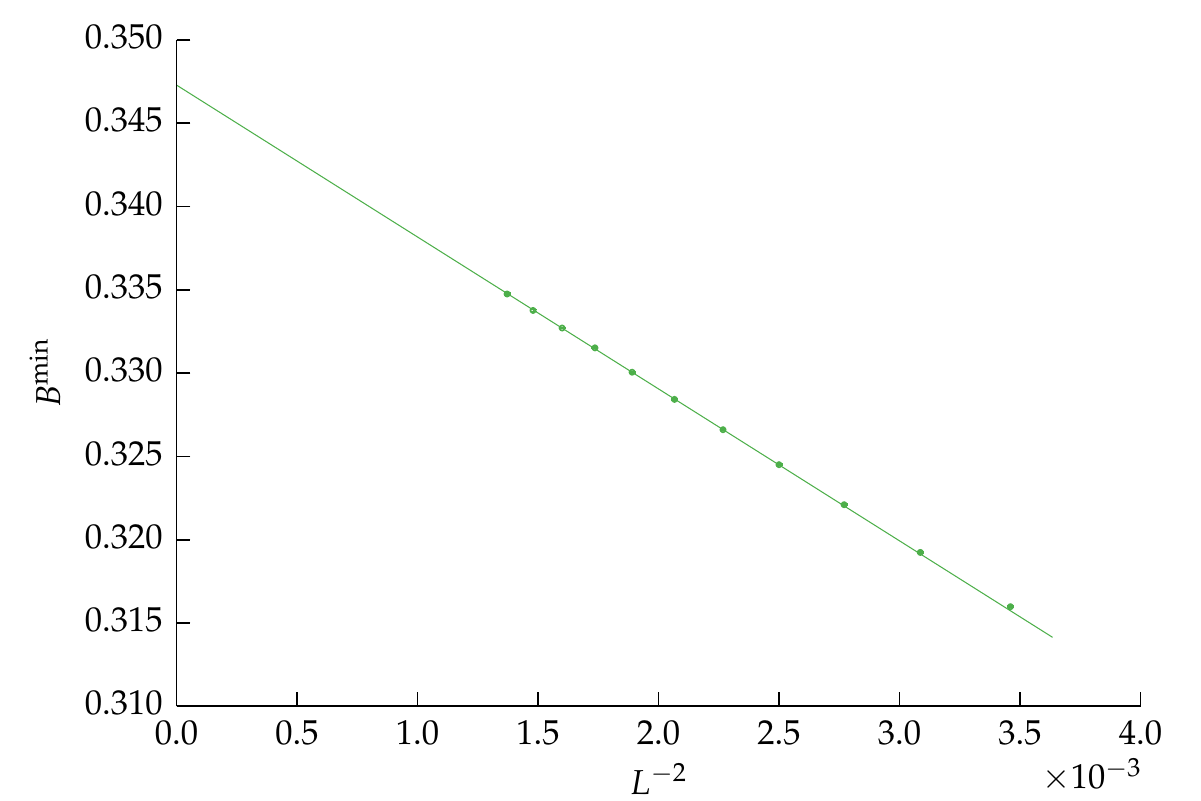}\vspace{1.0ex} 
		\caption{(a) Curves of the Binder parameter as function of $\beta$ and (b) their minima, ${B^{\rm
min}}(L)$, vs $1/L^2$ also showing the (non-standard) ${\cal O} (L^{-2})$ behaviour of the leading correction. The best fit line of $0.34729(7) - 9.12(4) L^{-2}$ through the measured values is drawn.} 
    \label{fig:ub_min_fit_nt} 
  \end{center} 
\end{figure} 

\begin{figure}[h] 
  \begin{center} 
    \includegraphics[scale=0.675]{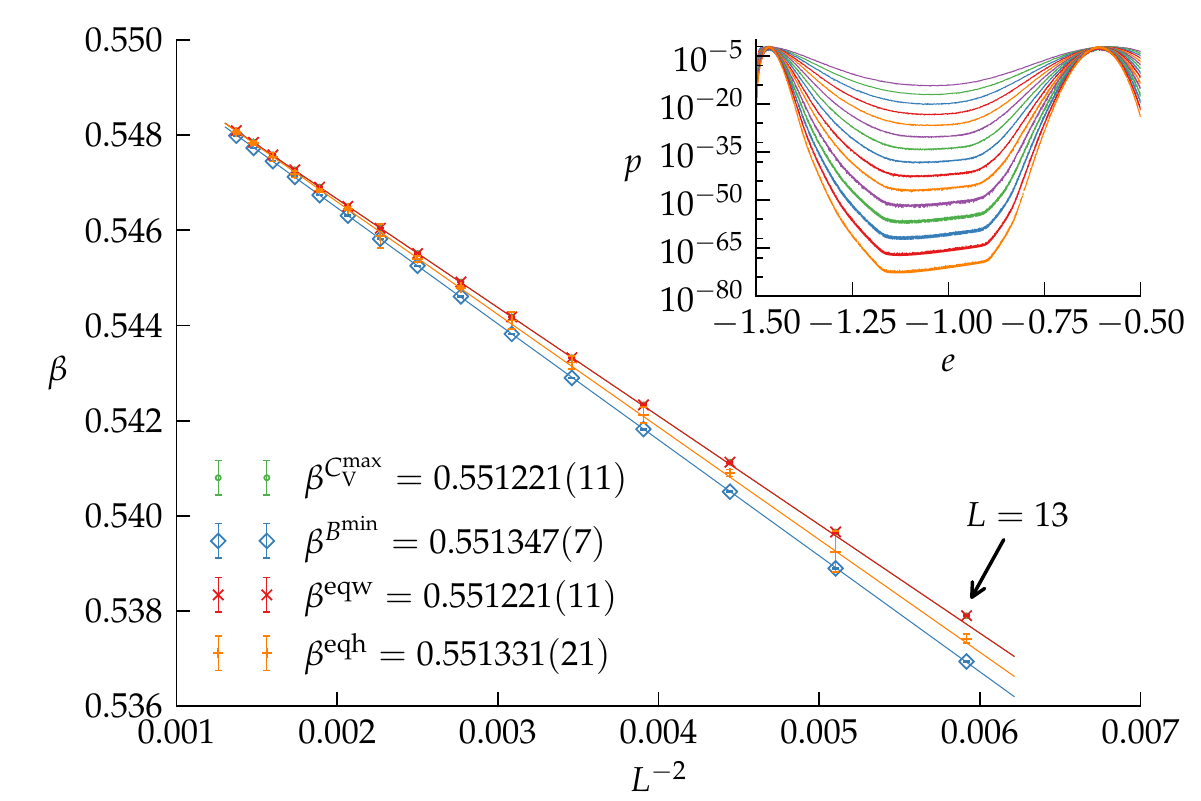} 
    \caption{
      Best fits using the leading $1/L^2$ scaling 
      for the (finite lattice) peak locations 
      of the specific
      heat $C_V^{\rm max}$, Binder's energy parameter $B^{\rm min}$; or inverse
      temperatures $\beta^{\rm eqw}$ and $\beta^{\rm eqh}$, where the two peaks
      of the energy probability density are of same weight or have equal
      height, respectively. The values for $\beta^{\rm eqw}$ and $\beta^{C_V^{\rm
      max}}$ are indistinguishable in the plot. The 
      omitted
      corrections which we discuss in detail
      in \cite{highero} give the slightly different effective slopes.
      The inset shows the energy
      probability density $p(e)$ over $e = E/L^d$ at $\beta^{\rm eqh}$ for
      lattice
      sizes
      $L\in \left\{ 13, 14, \dots, 26, 27 \right\}$. }
    \label{fig:fit:orig-pbc} 
  \end{center} 
\end{figure} 

We first verify the modified scaling for 
$C_V^{\rm max}(L)$ and $ B^{\rm min}(L)$ given in Eqs.~(\ref{eq:fss:specheat}), (\ref{eq:fss:binder}).  From 
 Figs.~\ref{fig:cv_max_fit_nt} and \ref{fig:ub_min_fit_nt} it is apparent that the $1/L^2$ scaling of the leading correction term engendered by the degeneracy $q=2^{3L}$ is clearly displayed by both quantities.
 The data and fits in Fig.~\ref{fig:fit:orig-pbc} for the two estimates for the inverse transition temperatures using Eqs.~(\ref{eq:fss:beta:specheat2}), (\ref{eq:fss:beta:binder2}) 
also clearly display the transmuted scaling laws with
$1/L^2$ corrections. In obtaining the best fit lines we have
left out the smaller lattices systematically, until a goodness-of-fit value of
at least $Q = 0.5$ was found for each observable individually.  
We have also included estimates of the transition temperatures 
using the additional estimators $\beta^{\rm eqw}(L)$
and $\beta^{\rm eqh}(L)$,
chosen
systematically to minimize 
\begin{equation} 
  D^{\rm eqw}(\beta) = \left( \sum_{e < e_{\rm min}} p(e, \beta) - \sum_{e \geq
  e_{\rm min}} p(e, \beta) \right)^2 \quad {\rm and} \quad 
  D^{\rm eqh}(\beta) =  \left( \max_{e < e_{\rm min}}\{p(e, \beta)\} - \max_{e
		    \geq e_{\rm min}}\{p(e, \beta)\} \right)^2 \; ,
\end{equation}
respectively, where the energy of the minimum between the two peaks, $e_{\rm min}$, is
determined beforehand to distinguish between the different phases.
%
From error weighted averages (refraining from a full cross-correlation analysis as discussed by \cite{combinefitsweigel2})
of the inverse transition temperatures $\beta^{C_V^{\rm max}}, 
\beta^{B^{\rm min}}, \beta^{\rm eqw}$, and $\beta^{\rm eqh}$ given in 
Fig.~\ref{fig:fit:orig-pbc}  we find
$\beta^{\infty} = {0.551\,291(7)}$
for the infinite lattice inverse transition temperature, where
the final error estimate is taken as the smallest error bar of the contributing $\beta$ estimates. 
The precision of the simulation results and the broad range of lattice
sizes clearly excludes fits in all cases to the standard finite-size scaling ansatz,
where the first correction is proportional to the inverse volume.

Any model with an exponentially degenerate low-temperature phase will display the modified scaling at a 
first-order phase transition described for the $3d$ gonihedric model here.
Apart from higher-dimensional variants of the gonihedric model or its dual, there are numerous other fields where the scenario could be realized. 
Examples range from ANNNI models discussed by \cite{selke} to topological {``orbital''} models in the context of quantum computing reviewed by \cite{nussinov} which all share an extensive ground-state degeneracy. Among the orbital models for transition metal compounds, a particularly promising candidate is the $3d$ classical compass or $t_{2g}$ orbital model   where a highly degenerate ground state is well known and signatures of a first-order transition into the disordered phase have recently been found numerically by \cite{compass4}.

Numerous other systems, such as the Ising antiferromagnet on a $3d$ FCC lattice,  have an exponentially
degenerate number of ground states but a small number of true low-temperature phases. Nonetheless, they do possess
an exponentially degenerate  number of low-energy excitations so, depending on the nature of the growth of energy barriers with system size, an {\it effective} modified scaling could still be seen at a first-order transition for the lattice sizes accessible in typical simulations. The crossover to the true asymptotic standard scaling would then only appear for very large lattices.
Indeed, previous simulations of \cite{beath_ryan} appear to have found non-standard scaling for the first-order transition in the Ising antiferromagnet on a $3d$ FCC lattice.

\vspace*{-2.4pt}
\section*{Acknowledgements}
This work was supported by
the Deutsch-Franz\"osische Hochschule (DFH-UFA) under Grant No.\ CDFA-02-07.


\vspace*{-2.4pt}
\begin{thebibliography}{100}

\bibitem[Beath and Ryan(2006)]{beath_ryan} Beath, A.D., Ryan, D.H., 2006.
Phys. Rev. B {73}, 174416.

\bibitem[Berg and Neuhaus(1991)]{muca1}
Berg, B.A., Neuhaus, T., 1991.
Phys.~Lett.~B {267}, 249.
%
\bibitem[Berg and Neuhaus(1992)]{muca2}
Berg, B.A., Neuhaus, T., 1992.
Phys.~Rev.~Lett.~{68}, 9.
    


\bibitem[Binder(1981)]{pioneer2}
Binder, K., 1981. Z. Phys. B {43}, 119.

\bibitem[Binder and Landau(1984)]{furtherd2}
Binder, K., Landau, D.P., 1984. Phys. Rev. B {30}, 1477.


\bibitem[Borgs and Janke(1992)]{borgs-janke}
Borgs, C., Janke, W., 1992.
Phys.~Rev.~Lett.~{68}, 1738.

\bibitem[Borgs and Koteck\'{y}(1990)]{rigorous1} 
Borgs, C., Koteck\'{y}, R., 1990.
J.~Stat.~Phys.~{61}, 79.
%


\bibitem[Borgs and Koteck\'{y}(1992)]{rigorous2} 
Borgs, C., Koteck\'{y}, R.,
1992. Phys. Rev. Lett. {68}, 1734.




\bibitem[Challa et al.(1986)]{furtherd3}
Challa, M.S.S., Landau, D.P., Binder, K., 1986. Phys. Rev. B {34}, 1841.


\bibitem[Espriu et al.(1997)]{firstorder}
  Espriu, D., Baig, M., Johnston, D.A., Malmini, R.K.P.C., 1997.
    J.~Phys.~A:~Math.~Gen.~{30}, 405.


\bibitem[Fisher and Berker(1982)]{pioneer3}
Fisher, M.E., Berker, A.N., 1982. Phys. Rev. B {26}, 2507.

\bibitem[Gerlach and Janke(2014)]{compass4} 
Gerlach, M.H., Janke, W., 2014.
First-order directional ordering transition in the 
three-dimensional compass model.
To be published.

\bibitem[Herrmann et al.(1992)]{1st-order-book}
Herrmann, H.J., Janke, W., Karsch, F. (Eds.), 1992.
``{\em Dynamics of First Order Phase Transitions\/}''.
World Scientific, Singapore.



\bibitem[Imry(1980)]{pioneer1}
Imry, Y., 1980. Phys. Rev. B {21}, 2042.




\bibitem[Janke(1992)]{mucaweights1}
Janke, W., 1992.
  Int. J. Mod. Phys. C {3}, 1137.
  
\bibitem[Janke(1993)]{twophasemodel1}
Janke, W., 1993.
    Phys.~Rev.~B {47}, 14757.
%
  
  
\bibitem[Janke(1998)]{mucaweights2}
Janke, W.,  
1998. Physica A {254}, 164.


\bibitem[Janke(2003a)]{janke_review}
Janke, W., 2003a.
In: ``{\em Computer Simulations of Surfaces and Interfaces\/}''.
D\"unweg, B., Landau, D.P., Milchev, A.I. (Eds.).
NATO Science Series, II. Math.,~Phys. and Chem. {114}, pp.\ 111--135, and references therein.

\bibitem[Janke(2003b)]{mucarecursion}
Janke, W., 2003b.
In: ``{\em Computer Simulations of Surfaces and Interfaces\/}''.
D\"unweg, B., Landau, D.P., Milchev, A.I. (Eds.).
NATO Science Series, II. Math.,~Phys. and Chem. {114}, pp.\ 137--157.

\bibitem[Johnston et al.(2008)]{goni_review}
Johnston, D.A., Lipowski, A., Ranasinghe, P.K.C.M., 2008. 
In: ``{\em Rugged Free-Energy Landscapes -- An Introduction}''. 
Janke, W. (Ed.). 
Springer Lecture Notes in Physics 736, pp.\ 173--199, and references therein.   

\bibitem[Lipowski(1997)]{glassy1}
Lipowski, A., 1997. J. Phys. A: Math.~Gen. {30}, 7365.

\bibitem[Mueller et al.(2014)]{highero}
Mueller, M., Janke, W., Johnston, D.A., 2014.
Phys. Rev. Lett. {112}, 200601.

\bibitem[Nussinov and van~den~Brink(2013)]{nussinov} 
Nussinov, Z., van~den~Brink, J., 2013.
Compass and Kitaev models -- Theory and physical motivations.
e-print {\tt arXiv:1303.5922}. 


\bibitem[Peczak and Landau(1989)]{furtherd4}
Peczak, P., Landau, D.P., 1989. Phys. Rev. B {39}, 11932.

\bibitem[Pietig and Wegner(1996)]{degen} 

Pietig, R., Wegner, F., 1996. Nucl. Phys. B {466}, 513.
%

\bibitem[Pietig and Wegner(1998)]{degen2} 
Pietig, R., Wegner, F., 1998. Nucl. Phys. B {525}, 549. 


\bibitem[Privman and Fisher(1983)]{furtherd1}
Privman, V., Fisher, M.E., 1983. J. Stat. Phys. {33}, 385.


\bibitem[Privman and Rudnik(1990)]{furtherd5}
Privman, V., Rudnik, J., 1990. J. Stat. Phys. {60}, 551.


\bibitem[Selke(1988)]{selke} 
Selke, W., 1988. Phys.~Rep. {170}, 213.



\bibitem[Weigel and Janke(2010)]{combinefitsweigel2}
Weigel, M., Janke, W., 2010.
Phys.~Rev.~E {81}, 066701.   



\end{thebibliography}
\end{document}